\documentclass[prc,superscriptaddress,unsortedaddress,
twocolumn,showpacs,preprintnumbers,amsmath,amssymb]{revtex4}

\usepackage{graphicx}
\usepackage{dcolumn}
\usepackage{bm}

\def\non{\nonumber}

\def\bfm#1{\mbox{\boldmath $#1$}}

\def\tfrac#1#2{\textstyle \frac{#1}{#2}}
\def\dder#1#2{\frac{d #1}{d #2}}
\def\d2der#1#2#3{\frac{d^2 #1}{d #2 d #3}}

\def\p2der#1#2#3{\frac{\partial^2 #1}{\partial #2 \partial #3}}

\begin{document}

\title{Analysis of $(K^-,K^+)$ inclusive spectrum with semiclassical distorted wave model}

\author{S. Hashimoto}
\affiliation{Department of Physics, Kyushu University, 
Fukuoka 812-8581, Japan}
\author{M. Kohno}
\affiliation{Physics Division, Kyushu Dental College, 
Kitakyushu 803-8580, Japan}
\author{K. Ogata}
\affiliation{Department of Physics, Kyushu University, 
Fukuoka 812-8581, Japan}
\author{M. Kawai}
\affiliation{Department of Physics, Kyushu University, 
Fukuoka 812-8581, Japan}

\date{\today}

\begin{abstract}
The inclusive $K^+$ momentum spectrum in the $^{12}$C$(K^-,K^+)$ 
reaction is calculated by the semiclassical distorted wave (SCDW) model, 
including the transition to the $\Xi^-$ bound state. The calculated 
spectra with the strength of the $\Xi^-$-nucleus potential $-50$, $-20$, 
and $+10$ MeV are compared with the experimental data measured at KEK 
with $p_{K^-}=1.65$ ${\rm GeV}/c$. The shape of the spectrum is 
reproduced by the calculation. Though the inclusive spectrum changes 
systematically depending on the potential strength, it is not possible 
to obtain a constraint on the potential from the present data. The 
calculated spectrum is found to have strong emission-angle dependence. 
We also investigate the incident $K^-$ momentum dependence of the 
spectrum to see the effect of the Fermi motion of the target nucleons 
which is explicitly treated in the SCDW method. 
\end{abstract}


\maketitle

\section{\label{sec-1}INTRODUCTION}
Hypernuclear physics has been disclosing characteristic features of 
strong interactions in baryon systems with strangeness. The 
$\Lambda$-$N$ interaction has been fairly well studied through analyses 
of experimental data on from light to heavy $\Lambda$ hypernuclei. 
$\Sigma$-$N$ and $\Xi$-$N$ interactions, however, are not well explored 
because of insufficient experimental data. In particular, the $\Xi$-$N$ 
interaction is much ambiguous for this reason despite the fact that its 
knowledge is the most important for understanding the physics of octet 
baryons as a probe of the strangeness $S=-2$ sector. 

Since the information from direct $Y$-$N$ scattering experiments is 
extremely limited, the data on the production of a hyperon $Y$ in 
strangeness exchange reactions $(\pi^+,K^+)$,~$(\pi^-,K^+)$,~$(K^-,K^+)$ 
etc. have been a principal source of information on the single-particle 
potential $U_Y$ felt by $Y$ in the final hypernucleus that reflects the 
properties of the underlying $Y$-$N$ interaction. The well depth of the 
$\Xi$-nucleus potential $U_{\Xi}$ was conjectured to be about 14 MeV by 
a distorted wave impulse approximation (DWIA) analysis of the inclusive 
$K^+$ spectra in the threshold region of the $(K^-,K^+)$ reaction 
\cite{Fukuda, Khaustov}. This potential depth is not conclusive, 
however, because of the limited statistics and resolution. The situation 
will be improved in the near future, since the $(K^-,K^+)$ experiments 
with high-intensity $K^-$ beams with a better resolution are prepared in 
the J-PARC project at KEK. 

In the recent papers \cite{Kohno,KFWOK}, the inclusive spectra of the 
$(\pi^-,K^+)$ $\Sigma$ formation reaction were analysed by means of the 
semiclassical distorted wave (SCDW) model \cite{Luo} and information on 
the $\Sigma$-nucleus potential was extracted. In this paper, we 
calculate, using SCDW, the inclusive $K^+$ spectrum in the $(K^-,K^+)$ 
$\Xi^-$ production reaction in an attempt to determine the 
$\Xi^-$-nucleus potential. In view of the on-going experimental 
progress, we try to elucidate the angle dependence and the incident 
energy dependence of the inclusive cross section. The SCDW model has 
been successful in quantitatively describing $(p,p^{\prime}x)$ and 
$(p,nx)$ inclusive cross sections for wide range of energy transfers and 
emission angles without any free adjustable parameter 
\cite{Luo,Watanabe,Ogata}. The single free parameter in the present SCDW 
calculation of the strangeness exchange reactions is the strength of the 
$\Xi^-$-nucleus potential $U_Y$. 

The formulation of the SCDW model for describing the $(K^-,K^+)$ 
spectrum is described in Sec. \ref{sec-2}: for quasifree region in 
Sec. \ref{sec-2-1} and for bound state region in Sec. \ref{sec-2-2}. In 
Sec. \ref{sec-3}, descriptions are given of the kaon distorted waves, 
the single-particle states of target nucleons and the $\Xi^-$ hyperon, 
and the strength of the elementary process $K^-+p\to K^++\Xi^-$. 
Numerical results of the calculations for the $(K^-,K^+)$ spectra are 
compared with the experimental data and discussed in Sec. \ref{sec-4}. A 
summary and conclusions are given in Sec. \ref{sec-5}. 

\section{\label{sec-2}FORMALISM}

\subsection{\label{sec-2-1}SCDW for inclusive spectrum}
The inclusive double differential cross section in the center of mass 
(c.m.) system for the $^{12}$C$(K^-,K^+)$ reaction with production of a 
$\Xi^-$ hyperon is given in the first order DWIA by 
\begin{eqnarray}
 \d2der{\sigma}{E_f}{\Omega_f}
  &=&
  (2\pi)^4 E_{i,{\rm red}}E_{f,{\rm red}} \frac{p_f}{p_i}
  \sum_{\alpha ,\, \beta}
  \frac{1}{4E_iE_f(2\pi)^6}
  \non \\
 && \times
  \left| \langle
  \chi_f^{(-)*}(\bfm{r}) \phi_{\beta}^*(\bfm{r})
  \mid v \mid
  \chi_i^{(+)}(\bfm{r}) \phi_{\alpha}(\bfm{r})
  \rangle \right|^2
  \non \\
 && \times
  \delta(\varepsilon_{\beta} - \varepsilon_{\alpha} - \omega)
  \,,
\label{eq-DDX1}
\end{eqnarray}
where $E_{c,{\rm red}}=E_cE_{c,A}/(E_c+E_{c,A})$ is the reduced energy, 
$p_c$ is the relative momentum between the kaon and the target $^{12}$C 
nucleus, $\chi_c^{(+) {\rm or} (-)}$ is the distorted wave of the kaon 
at energy $E_c$ for $c=i$ and $f$ where $i$ and $f$ stand for the 
initial and the final state, respectively, and $E_{i,A}$ ($E_{f,A}$) is 
the total energy of the target nucleus (hypernucleus) in the c.m. 
system. The factor $4E_iE_f(2\pi)^6$ is the product of the normalization 
factors of the $\chi_c$ given by 
\begin{eqnarray}
 \langle
  \chi_c(\bfm{p}_c)|\chi_c(\bfm{p}_c^{\prime})
  \rangle
  =
  2E_c(2\pi)^3\delta^3(\bfm{p}_c-\bfm{p}_c^{\prime})
  \,.
\end{eqnarray}
The suffices $\alpha$ and $\beta$ denote the states of the nucleon hole 
and the unobserved $\Xi^-$ hyperon in the final hypernucleus, 
respectively, and $\phi_{\alpha}$ and $\phi_{\beta}$ are the 
corresponding single particle wave functions with energies 
$\varepsilon_{\alpha}$ and  $\varepsilon_{\beta}$. The delta function 
guarantees energy conservation where $\omega=E_i-E_f$ is the energy 
transfer. The $t$-matrix in coordinate representation of the elementary 
process $K^- + p \to K^+ + \Xi^-$ is denoted by $v$ that depends on 
energies and momenta of the particles in the reaction. In the present 
calculation we assume the range of $v$ to be zero. 

In the laboratory system, in which most experimental data are given, the 
cross section is given by
\begin{eqnarray}
 \d2der{\sigma}{p_f^{\rm L}}{\Omega_f^{\rm L}}
  =
  J
  \d2der{\sigma}{E_f}{\Omega_f}
  \,,
\end{eqnarray}
where ${\rm L}$ denotes the laboratory system and 
$J=\partial(E_f,\Omega_f)/\partial(p_f^{\rm L},\Omega_f^{\rm L})$ 
is the Jacobian. 

Expanding the squared modulus in Eq. (\ref{eq-DDX1}), one obtains 
\begin{widetext}
\begin{eqnarray}
 \d2der{\sigma}{p_f^{\rm L}}{\Omega_f^{\rm L}}
  =
  J
  (2\pi)^4 E_{i,{\rm red}}E_{f,{\rm red}} \frac{p_f}{p_i}
  \frac{1}{4E_iE_f(2\pi)^6}
  \int \int d\bfm{r} d\bfm{r}^{\prime} \,
  \chi_f^{(-)*}(\bfm{r})
  v
  \chi_i^{(+)}(\bfm{r})
  \chi_f^{(-)}(\bfm{r}^{\prime})
  v^*
  \chi_i^{(+)*}(\bfm{r}^{\prime})
  K(\bfm{r},\bfm{r}^{\prime})
  \,,
\label{eq-DDX2}
\end{eqnarray}
\end{widetext}
where the nonlocal kernel $K(\bfm{r},\bfm{r}^{\prime})$ is given by 
\begin{eqnarray}
 K(\bfm{r},\bfm{r}^{\prime})
  &=&
  \sum_{\alpha ,\, \beta}
  \phi_{\beta}^*(\bfm{r}) \phi_{\alpha}(\bfm{r})
  \phi_{\beta}(\bfm{r}^{\prime}) \phi_{\alpha}^*(\bfm{r}^{\prime})
  \non \\
 && \times
  \delta(\varepsilon_{\beta} - \varepsilon_{\alpha} - \omega)
  \,.
\label{eq-kernel1}
\end{eqnarray}
The summation over $\alpha$ extends to all the hole states below the 
Fermi level. The summation over $\beta$ is unrestricted since there is 
no Pauli principle for the produced hyperon. 

For the summation over $\alpha$, we use the Wigner transform 
\begin{eqnarray}
 \sum_{\alpha} \phi_{\alpha}(\bfm{r}) \phi_{\alpha}^*(\bfm{r}^{\prime})
  =
  \sum_{nlj}
  \int d\bfm{k}_{\alpha} \,
  \Phi_{nlj}(\bfm{R}, \bfm{k}_{\alpha})
  e^{-i\bfm{k}_{\alpha}\cdot\bfm{s}}
  \,,
\end{eqnarray}
where $\bfm{R}=\frac{\bfm{r}+\bfm{r}^{\prime}}{2}$ and 
$\bfm{s}=\bfm{r}^{\prime}-\bfm{r}$ denote the midpoint and the relative 
coordinates of $\bfm{r}$ and $\bfm{r}^{\prime}$, respectively. The 
summation over $\beta$ is replaced by the integrals 
\begin{eqnarray}
 \sum_{\beta}
  \phi_{\beta}^*(\bfm{r}) \phi_{\beta}(\bfm{r}^{\prime})
  &\to&
  \frac{1}{(2 \pi)^3}
  \int d\Omega_{\beta}
  \int d\varepsilon_{\beta}
  \frac{1}{2}
  \left(\frac{2\mu_Y}{\hbar^2}\right)^{3/2}
  \non \\
 && \times
  \sqrt{\varepsilon_{\beta}}\,\,\,
  \phi_{\beta}^*(\bfm{r}) \phi_{\beta}(\bfm{r}^{\prime})
  \,,
\end{eqnarray}
over the outgoing direction $\Omega_{\beta}$ and the energy 
$\varepsilon_{\beta}$ of the produced hyperon in state $\beta$ in the 
asymptotic region, and $\mu_Y$ is the reduced mass of the 
$\Xi^-$-nucleus system in the final state. 

In the SCDW model, the local semiclassical approximation (LSCA) is used 
for the distorted waves $\chi_c$ \cite{Luo} 
\begin{eqnarray}
 \chi_c(\bfm{R}\pm\tfrac{1}{2}\bfm{s})
  \cong
  \chi_c(\bfm{R})
  e^{\pm i \bfm{k}_c \bigl(\bfm{R}\bigr) \cdot \tfrac{1}{2}\bfm{s}}
  \,,
\label{eq-LSCA}
\end{eqnarray}
for small $\bfm{s}/2$, where $\hbar\bfm{k}_c(\bfm{R})$ is the local 
momentum. The direction of $\bfm{k}_c(\bfm{R})$ is calculated by 
\begin{eqnarray}
 \hat{\bfm{k}}_c(\bfm{R})
  =
  \frac{\rm{Re}[\chi_c^{(\pm)*}(\bfm{R})
  (-i)\nabla\chi_c^{(\pm)}(\bfm{R})]}
  {\left|\chi_c^{(\pm)}(\bfm{R})\right|^2}
  \,,
\label{eq-LMD}
\end{eqnarray}
and the magnitude is determined by the energy-momentum relation at 
$\bfm{R}$, 
\begin{eqnarray}
 \hbar^2 c^2k_c^2(\bfm{R})
  + 2E_{c,{\rm red}}(U_R(\bfm{R})+V_{\rm Coul}(\bfm{R}))
  \non \\
 - V_{\rm Coul}^2(\bfm{R})
 =
 E_c^2-m_K^2c^4
 \,,
\end{eqnarray}
where $m_K$ is the kaon mass 494 ${\rm MeV}/c^2$, 
$V_{\rm Coul}(\bfm{R})$ is the Coulomb potential, and $U_R(\bfm{R})$ is 
the real part of the distorting potential for $\chi_c$ with $E_c$. 

The LSCA for the distorted waves $\chi_c$ is expected to work well, 
because the nonlocal kernel $K(\bfm{r},\bfm{r}^{\prime})$, which can now 
be written as 
\begin{eqnarray}
 K(\bfm{R}-\tfrac{1}{2}\bfm{s}, \bfm{R}+\tfrac{1}{2}\bfm{s})
  &=&
  \frac{1}{2(2 \pi)^3}
  \left(\frac{2\mu_Y}{\hbar^2}\right)^{3/2}
  \sum_{nlj}
  \int d\bfm{k}_{\alpha} \,
  \non \\
 && \times
  \Phi_{nlj}(\bfm{R}, \bfm{k}_{\alpha})
  e^{-i\bfm{k}_{\alpha}\cdot\bfm{s}}
  \int d\Omega_{\beta}
  \non \\
 && \times
  \int d\varepsilon_{\beta}
  \sqrt{\varepsilon_{\beta}}
  \phi_{\beta}^*(\bfm{R}-\tfrac{1}{2}\bfm{s})
  \non \\
 && \times
  \phi_{\beta}(\bfm{R}+\tfrac{1}{2}\bfm{s})
  \delta(\varepsilon_{\beta} - \varepsilon_{\alpha} - \omega)
  \,,
  \non \\
\end{eqnarray}
is appreciable only for small $\bfm{s}$ because of the completeness 
$\sum_{\beta}\phi_{\beta}^*(\bfm{R}-\bfm{s}/2)\phi_{\beta}(\bfm{R}+\bfm{s}/2)=\delta(\bfm{s})$. 
It is noted that the range of the kernel is shorter than in the cases of 
$(p,p^{\prime}x)$ and $(p,nx)$, in which the Pauli principle restricts 
the summation over $\beta$. 

In the present calculation, we make use of the LSCA for the unbound wave 
functions $\phi_{\beta}$ in the $\Xi^-$-nucleus potential $U_{\Xi}$, 
hence, 
$\phi_{\beta}(\bfm{R}\pm\bfm{s}/2)\cong\phi_{\beta}(\bfm{R})\exp[\pm i \bfm{k}_{\beta}\bigl(\bfm{R}\bigr)\cdot\bfm{s}/2]$
with the direction given from the flux of $\phi_{\beta}(\bfm{R})$ as in 
Eq. (\ref{eq-LMD}) and the magnitude 
$k_{\beta}\bigl(\bfm{R}\bigr)=\{2\mu_Y[\varepsilon_{\beta}-U_{\Xi}(\bfm{R})]/\hbar^2\}^{1/2}$, 
where $U_{\Xi}(\bfm{R})$ is assumed to be real as described in 
Sec. \ref{sec-3}. 
The approximation is valid when $\varepsilon_{\beta}$ is large compared 
with $U_{\Xi}(\bfm{R})$ and $\bfm{k}_{\beta}\bigl(\bfm{R}\bigr)$ is a 
slowly function of $\bfm{R}$. 

Using the approximations described above, the double differential cross 
section for the inclusive reaction becomes 
\begin{widetext}
\begin{eqnarray}
 \d2der{\sigma}{p_f^{\rm L}}{\Omega_f^{\rm L}}
  &=&
  J
  \frac{E_{i,{\rm red}}E_{f,{\rm red}}}{2(2\pi)^2} \frac{p_f}{p_i}
  \left(\frac{2\mu_Y}{\hbar^2}\right)^{3/2}
  \int d\bfm{R} \,
  \sum_{nlj}
  \frac{1}{4E_iE_f}
  \int d\bfm{k}_{\alpha}
  \int d\Omega_{\beta}
  \int d\varepsilon_{\beta}
  \sqrt{\varepsilon_{\beta}}
  \left|\chi_f^{(-)}(\bfm{R})\right|^2
  \left|\chi_i^{(+)}(\bfm{R})\right|^2
  \left|v\right|^2
  \non \\
 && \times
  \Phi_{nlj}(\bfm{R}, \bfm{k}_{\alpha})
  \left|\phi_{\beta}(\bfm{R})\right|^2
  \delta\Bigl(\bfm{k}_f\bigl(\bfm{R}\bigr) - \bfm{k}_i\bigl(\bfm{R}\bigr)
  + \bfm{k}_{\beta}\bigl(\bfm{R}\bigr) - \bfm{k}_{\alpha}\Bigr)
  \delta(\varepsilon_{\beta} - \varepsilon_{\alpha} - \omega)
  \,,
\label{eq-DDX3}
\end{eqnarray}
where the first delta function on the right hand side implies the local 
momentum conservation. 

\subsection{\label{sec-2-2}SCDW for exclusive spectrum}
We apply the SCDW model also to the $(K^-,K^+)$ spectrum in the region 
of bound states of hypernucleus. The differential cross section in the 
laboratory system for the transition to a $\Xi^-$-$^{11}$B bound state 
is given by 
\begin{eqnarray}
 \dder{\sigma}{\Omega_f^{\rm L}}
  &=&
  J^{\prime}
  (2\pi)^4 E_{i,{\rm red}}E_{f,{\rm red}} \frac{p_f}{p_i}
  \int \int d\bfm{r} d\bfm{r}^{\prime} \,
  \frac{1}{4E_iE_f(2\pi)^6}
  \chi_f^{(-)*}(\bfm{r}) \phi_{\beta^{\prime}}^*(\bfm{r})
  v_{f \beta^{\prime} ,\, i \alpha^{\prime}}
  \chi_i^{(+)}(\bfm{r}) \phi_{\alpha^{\prime}}(\bfm{r})
  \non \\
 && \times
  \chi_f^{(-)}(\bfm{r}^{\prime}) \phi_{\beta^{\prime}}(\bfm{r}^{\prime})
  v_{f \beta^{\prime} ,\, i \alpha^{\prime}}^*
  \chi_i^{(+)*}(\bfm{r}^{\prime}) \phi_{\alpha^{\prime}}^*(\bfm{r}^{\prime})
  \,,
\label{eq-DXB1}
\end{eqnarray}
where $J^{\prime}=\partial\Omega_f/\partial\Omega_f^{\rm L}$ is the 
Jacobian. Compared with the inclusive spectrum, this formula has neither 
the energy delta function nor the summation of $\alpha$ or $\beta$. 
Corresponding to a transition from proton-hole state $\alpha^{\prime}$ 
to a $\Xi^-$-particle state $\beta^{\prime}$ with the energies 
$\varepsilon_{\alpha^{\prime}}$ and $\varepsilon_{\beta^{\prime}}$, 
respectively, there should be a spike at the point with the energy 
transfer 
$\omega=\varepsilon_{\beta^{\prime}}-\varepsilon_{\alpha^{\prime}}$. 

We use the Wigner transforms of the product of wave functions of the 
nucleon hole and those of the hyperon:
\begin{eqnarray}
 \phi_{\alpha^{\prime}}(\bfm{r}) \phi_{\alpha^{\prime}}^*(\bfm{r}^{\prime})
  &=&
  \int d\bfm{k}_{\alpha^{\prime}} \,
  \Phi_{\alpha^{\prime}}(\bfm{R}, \bfm{k}_{\alpha^{\prime}})
  e^{-i\bfm{k}_{\alpha^{\prime}}\cdot\bfm{s}}
  \,,
\label{eq-WigAp}
\\
 \phi_{\beta^{\prime}}^*(\bfm{r}) \phi_{\beta^{\prime}}(\bfm{r}^{\prime})
  &=&
  \int d\bfm{k}_{\beta^{\prime}} \,
  \Phi_{\beta^{\prime}}(\bfm{R}, \bfm{k}_{\beta^{\prime}})
  e^{i\bfm{k}_{\beta^{\prime}}\cdot\bfm{s}}
  \,.
\label{eq-WigBp}
\end{eqnarray}
Using Eqs. (\ref{eq-WigAp}) and (\ref{eq-WigBp}) and changing the 
variables of integration to ($\bfm{R}$,~$\bfm{s}$), one can write 
Eq. (\ref{eq-DXB1}) as 
\begin{eqnarray}
 \dder{\sigma}{\Omega_f^{\rm L}}
  &=&
  J^{\prime}
  \frac{E_{i,{\rm red}}E_{f,{\rm red}}}{(2\pi)^2} \frac{p_f}{p_i}
  \int d\bfm{R} \,
  \frac{1}{4E_iE_f}
  \int d\bfm{k}_{\alpha^{\prime}}
  \int d\bfm{k}_{\beta^{\prime}}
  \left|\chi_f^{(-)}(\bfm{R})\right|^2
  \left|\chi_i^{(+)}(\bfm{R})\right|^2
  \left|v\right|^2
  \non \\
 && \times
  (2\pi)^3
  \Phi_{\alpha^{\prime}}(\bfm{R}, \bfm{k}_{\alpha^{\prime}})
  \Phi_{\beta^{\prime}}(\bfm{R}, \bfm{k}_{\beta^{\prime}})
  \delta\Bigl(\bfm{k}_f\bigl(\bfm{R}\bigr) - \bfm{k}_i\bigl(\bfm{R}\bigr)
  + \bfm{k}_{\beta^{\prime}} - \bfm{k}_{\alpha^{\prime}}\Bigr)
  \,.
\label{eq-DXB2}
\end{eqnarray}
In this case, the nonlocal kernel 
$K(\bfm{r},\bfm{r}^{\prime})=\phi_{\alpha^{\prime}}(\bfm{r})\phi_{\alpha^{\prime}}^*(\bfm{r}^{\prime})\phi_{\beta^{\prime}}^*(\bfm{r})\phi_{\beta^{\prime}}(\bfm{r}^{\prime})$ 
has the range $\bfm{s}$ of the order of the target nuclear radius. Since 
we still use the LSCA for the $\chi_c$, the accuracy of 
Eq. (\ref{eq-DXB2}) depends on that of the LSCA with such large 
$\bfm{s}$, which will be discussed in Sec. \ref{sec-4-2}. 
\end{widetext}

\section{\label{sec-3}WAVE FUNCTIONS AND TRANSITION STRENGTH}
The distorted waves of the $K^{\pm}$ are calculated as solutions of the 
standard Klein-Gordon equation with distorting potentials $U_c$. The 
equation in the laboratory system reads 
\begin{eqnarray}
 \bigl[
  \hbar^2 c^2\nabla_{\bfm{r}}^2 + (p_K^{\rm L})^2c^2
  - 2E_K^{\rm L}(U_c^{\rm L}(\bfm{r})+V_{\rm Coul}^{\rm L}(\bfm{r}))
  \non \\
 + (V_{\rm Coul}^{\rm L}(\bfm{r}))^2
  \bigr]
  \chi_c^{\rm L}(\bfm{r})
  = 0
  \,,
\label{eq-KG1}
\end{eqnarray}
where $\bfm{r}$ is the displacement of the kaon from the center of mass 
of the target nucleus, $p_K^{\rm L}$ ($E_K^{\rm L}$) is the momentum 
(energy) of the kaon in the laboratory system, and $U_c^{\rm L}$ and 
$V_{\rm Coul}$ are, respectively, the distorting and the Coulomb 
potentials defined in the laboratory system for the states $c=i$ and 
$f$. In the present calculation, we approximate Eq. (\ref{eq-KG1}) by 
the equation in the c.m. system 
\begin{eqnarray}
 \bigl[
  \hbar^2 c^2\nabla_{\bfm{r}}^2 + p_c^2c^2
  - 2E_{c,{\rm red}}(U_c(\bfm{r})+V_{\rm Coul}(\bfm{r}))
  \non \\
 + V_{\rm Coul}^2(\bfm{r})
  \bigr]
  \chi_c(\bfm{r})
  = 0
  \,,
\label{eq-KG2}
\end{eqnarray}
replacing $p_K^{\rm L}$, $E_K^{\rm L}$, $U_c^{\rm L}$, and 
$V_{\rm Coul}$ by $p_c$, $E_{c,{\rm red}}$, $U_c$, and $V_{\rm Coul}$, 
respectively, in order to include the nuclear recoil effect. We use 
distorting potentials of the standard $t\rho$ approximation of 
Ref. \cite{Friedman}: 
\begin{eqnarray}
 2E_{c,{\rm red}}U_c(r)
  &=&
  -4\pi F_k
  \left(i\frac{\sigma_{KN}k_{KN}}{4\pi}+{\rm Re}[f_{KN}(0)]\right)
  \non \\
 && \times
  \rho(r)
  \,,
\end{eqnarray}
where $F_k=M_{c,A}W_{KN}/M(E_c+E_{c,A})$ is a kinematical factor 
resulting from the transformation of the $KN$ scattering amplitude from 
the $K$-$N$ to the $K$-nucleus c.m. systems, $\sigma_{KN}$ is the 
isospin-averaged $KN$ total cross section, $f_{KN}(0)$ is the $KN$ 
forward scattering amplitude in the free space, and $\rho(r)$ is the 
nucleon density distribution of the nucleus in the $K$-nucleus c.m. 
system. The ${\rm Re}[f_{KN}(0)]$ and $\sigma_{KN}$ are obtained as the 
isospin-average of those given in Ref. \cite{Sibirtsev2} in 
parameterized forms. The masses of the nucleon, the target nucleus, and 
the hypernucleus are denoted by $M$, $M_{i,A}$, and $M_{f,A}$, 
respectively, and $W_{KN}$ ($k_{KN}$) denotes the total kaon-nucleon 
energy (wave number) in their c.m. system. 

\begin{figure}
\includegraphics[width=70mm,keepaspectratio]{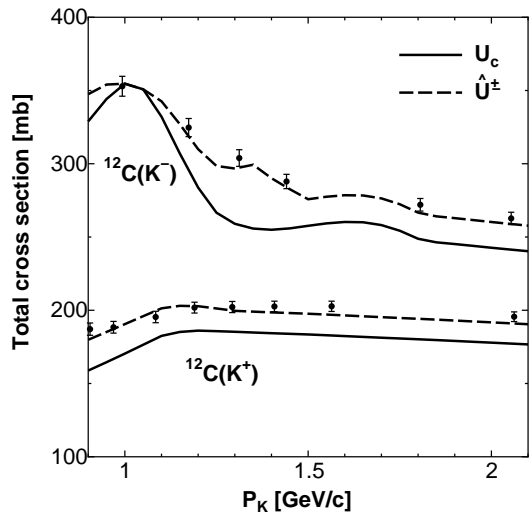}
\caption{\label{fig-12CKpmTX}
The total cross sections of the $K^-$ and the $K^+$ collisions on 
$^{12}$C as functions of the incident momenta of $K^-$ and $K^+$ 
in the laboratory frame. The optical model values with the $t\rho$ 
potentials $U_c$ and the optical potentials $\hat{U}^{\pm}$ are 
represented by the solid lines and the dashed lines, respectively, and 
compared with the experimental data of Ref. \cite{Bugg}. 
}
\end{figure}

Actually, the $U_c$ do not agree with the optical potentials 
$\hat{U}^{\pm}$ that fit the experimental total cross sections 
$\sigma_{\rm T}$ of $K^{\pm}$. Figure \ref{fig-12CKpmTX} shows the 
$\sigma_{\rm T}$ for $K^{\pm}$ on $^{12}$C calculated with the 
$\hat{U}^{\pm}$ (dashed line) and the $U_c$ (solid line). It turns out, 
however, that the inclusive $(K^-,K^+)$ cross section calculated with 
the $U_c$ agrees with that calculated with the $\hat{U}^{\pm}$ within 
10\% which is within the experimental error of the $(K^-,K^+)$ spectrum. 
We use, therefore, the potentials $U_c$ in our calculations. 

The single-particle wave functions and the density distribution 
function $\rho(r)$ of the target nucleus are calculated in the 
Hartree-Fock model of Campi and Sprung \cite{Campi}. The unobserved 
hyperon $\Xi^-$ is assumed to propagate in a Woods-Saxon potential, 
\begin{eqnarray}
 U_{\Xi}(r)
  =
  \frac{V_{\Xi}}{1+\exp\{(r-R)/a\}}
\end{eqnarray}
with $R=1.1A^{1/3}$ fm and $a=0.65$ fm. The calculation is performed 
with a series of $V_{\Xi}=-50$, $-20$, and $+10$ MeV. The flux of the 
$\Xi^-$ decreases in the propagation because of its interaction with 
other nucleons. We take that effect and also the experimental energy 
resolution into account by means of convolution of the calculated 
spectra with a Lorentz form of width $\Gamma$. 

The transition strength $|v|$ of the elementary process 
$K^-+p\to K^++\Xi^-$ is assumed to be related to the differential cross 
section in the free space 
\begin{eqnarray}
 \dder{\sigma}{\Omega}
  =
  \frac{1}{(4\pi)^2} \frac{E_pE_{\Xi}}{W^2} \frac{k_{K^+}}{k_{K^-}}
  \left|v\right|^2
\end{eqnarray}
in the on-shell approximation, where $W$ is the invariant energy, and 
$E_p$ ($E_{\Xi}$) is the energy of the proton (the $\Xi^-$ hyperon) and 
$k_{K^-}$ ($k_{K^+}$) is the momentum of $K^-$ ($K^+$) in the $K^-$-$p$ 
($K^+$-$\Xi^-$) c.m. system. We use the elementary cross section 
parameterized in Ref. \cite{Nara} as a function of $s$ and the 
scattering angle $\theta$. In our model described in the previous 
section, $W$ and $\theta$ are given by the local momenta of the 
particles $K^-$ and $p$ in the initial state and $K^+$ and $\Xi^-$ in 
the final state at each collision point $\bfm{R}$. 

\section{\label{sec-4}RESULTS AND DISCUSSION}

\subsection{\label{sec-4-1}Transition to continuum states}
\begin{figure}
\includegraphics[width=70mm,keepaspectratio]{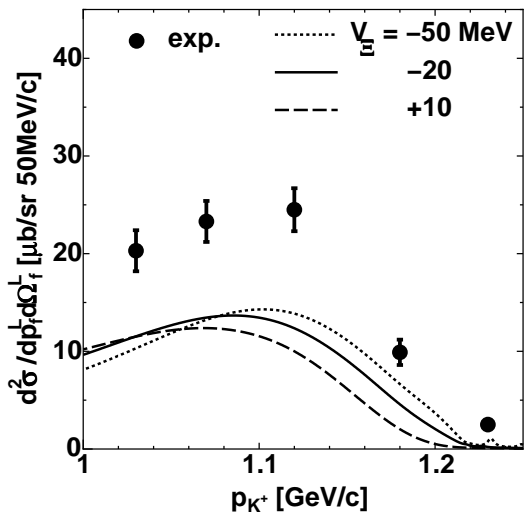}
\caption{\label{fig-12CKmKp-Vdep}
Momentum spectrum of $K^+$ in $^{12}$C$(K^-,K^+)$ at $p_{K^-}=1.65$ 
${\rm GeV}/c$ averaged over the emission angles between 
$\theta_{K^+}=1.7^\circ$ and $13.6^\circ$. The data of 
Ref. \cite{Iijima} are compared with theoretical values calculated with 
$V_{\Xi}=-50$, $-20$, and $+10$ MeV represented by the curves as 
indicated. 
}
\end{figure}

In Fig. \ref{fig-12CKmKp-Vdep} the momentum spectrum of $K^+$ in 
$^{12}$C$(K^-,K^+)$ at incident momentum $p_{K^-}=1.65$ ${\rm GeV}/c$ 
calculated with $V_{\Xi}=-50$ (dotted line), $-20$ (solid line), and 
$+10$ MeV (dashed line) are compared with the KEK experimental data 
\cite{Iijima}. Since the data are average over the emission angles 
$\theta_{K^+}$ between $1.7^{\circ}$ and $13.6^{\circ}$, the calculated 
values are correspondingly averaged. The spectrum shown is after the 
Lorentzian convolution mentioned in the preceding section. The result 
turns out to be quite insensitive to the width $\Gamma$ of the 
convolution. In Fig. \ref{fig-12CKmKp-Vdep} the half width $\Gamma/2$ is 
tentatively taken to be $2$ MeV, the same value as for the transitions 
to bound states discussed later. 

The shape of the spectrum is reproduced by the calculation, but the 
absolute magnitude of the cross section is underestimated by about a 
factor of 2. A possible reason on the theoretical side is that we do not 
correct the lack of the center-of-mass motion of the target wave 
function. Since the momentum transfer from $K^-$ to $K^+$ is large, the 
effect is not negligible. This problem in the context of the SCDW method 
has yet to be clarified. On the physics side, we ignore multi-step 
processes. Their contributions have been estimated by Nara {\it et al.} 
\cite{Nara} by means of the intranuclear cascade model (INC), and shown 
to be small in the case of $^{12}$C$(K^-,K^+)$. Since INC is a classical 
mechanical simulation, however, reinvestigation by means of the quantum 
mechanical SCDW model is desirable, even though the magnitude of the 
cross sections in Fig. \ref{fig-12CKmKp-Vdep} is almost the same as that 
of the one-step process in given by the INC calculation of 
Ref. \cite{Nara}. Another possible reason for the discrepancy is the 
in-medium modification of the elementary process which is not taken into 
account in the calculation. There is also experimental uncertainty of 
about 30 \% in the magnitude of the elementary cross section. 
Figure \ref{fig-12CKmKp-Vdep} shows that the $V_{\Xi}$-dependence of the 
angle-averaged spectrum is not very strong. The peak position shifts 
naturally as the potential strength changes. It is obvious, however, 
that one hardly determines the value of $V_{\Xi}$ from the present 
experimental data. 

\begin{figure}
\includegraphics[width=70mm,keepaspectratio]{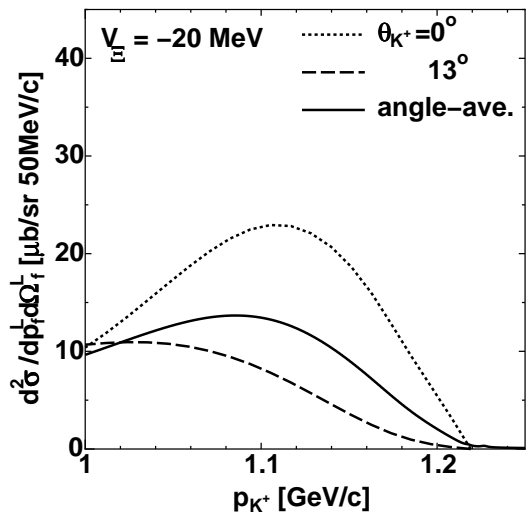}
\caption{\label{fig-12CKmKp-Adep}
Emission angle dependence of the momentum spectrum in $^{12}$C$(K^-,K^+)$ 
at $p_{K^-}=1.65$ ${\rm GeV}/c$ with $V_{\Xi}=-20$ MeV. The spectra at 
$\theta_{K^+}=0^\circ$ and $13^\circ$ are shown together with the 
average over $1.7^\circ < \theta_{K^+} < 13.6^\circ$. 
}
\end{figure}

Figure \ref{fig-12CKmKp-Adep} shows the same calculated momentum spectra 
as in Fig. \ref{fig-12CKmKp-Vdep} at $\theta_{K^+}=0^{\circ}$ (dotted 
line) and $13^{\circ}$ (dashed line). One sees that the $\theta_{K^+}$ 
dependence is very strong. This is in contrast to the almost same 
spectra at $0^{\circ}$ and $13.6^{\circ}$ in Ref. \cite{Tadokoro} 
obtained with angle-averaged elementary cross sections. This clearly 
shows the significance of the use of angular dependent elementary cross 
sections in the present SCDW model calculation. 

\begin{figure}
\includegraphics[width=70mm,keepaspectratio]{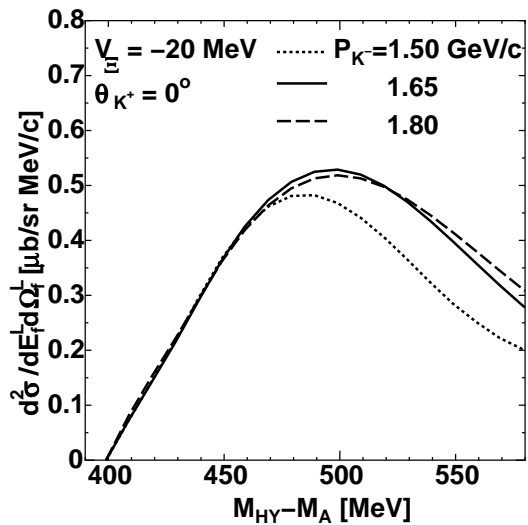}
\caption{\label{fig-12CKmKp-Edep}
The energy spectrum of $K^+$ at $\theta_{K^+}=0^\circ$ at 
$p_{K^-}=1.50$, $1.65$, and $1.80$ ${\rm GeV}/c$ calculated with 
$V_{\Xi}=-20$ MeV and $\Gamma/2=0$, are plotted as functions of 
$M_{HY}-M_A$ where $M_{HY}$ and $M_A$ are the masses of the hypernucleus 
and the target nucleus, respectively. 
}
\end{figure}

Figure \ref{fig-12CKmKp-Edep} shows the dependence of the calculated 
$^{12}$C$(K^-,K^+)$ energy spectrum on the incident momentum $p_{K^-}$, 
which is not very strong. The dotted, solid, and dashed curves denote 
the calculated energy spectra at $p_{K^-}=1.50$, $1.65$, and $1.80$ 
${\rm GeV}/c$, respectively. It is interesting to note, however, that 
the change of the position and the hight of the peak with the change of 
$p_{K^-}$ from $1.50$ to $1.80$ ${\rm GeV}/c$ is not simple, unlike that 
of the elementary cross section of Ref. \cite{Nara} used in the present 
calculation. Analysis of the calculation shows that the Fermi motion of 
the target nucleon is responsible for this somewhat complicated 
$p_{K^-}$ dependence of the spectrum. 

\subsection{\label{sec-4-2}Transition to bound states}
\begin{figure}
\includegraphics[width=70mm,keepaspectratio]{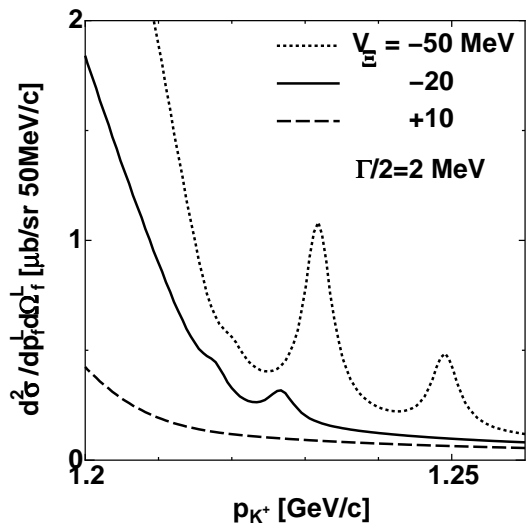}
\caption{\label{fig-12CKmKp-Vdep-th}
The $^{12}$C$(K^-,K^+)$ spectra near the $\Xi^-$-$^{11}$B hypernuclear 
bound state region at $p_{K^-}=1.65 {\rm GeV}/c$ and 
$\theta_{K^+}=8^{\circ}$ calculated with $V_{\Xi}=-50$, $-20$, and $+10$ 
MeV. 
}
\end{figure}

If the potential $U_{\Xi}$ is attractive and sufficiently strong, the 
hypernuclear system $\Xi^-$-$^{11}$B can have bound states. 
Table \ref{tab-EB} shows the energy eigenvalues of the single-particle 
states of $\Xi^-$ in $U_{\Xi}$ for the depth of $V_{\Xi}=-20$ and $-50$ 
MeV. Transitions to states with a nucleon hole in the 
$0s_{1/2}^{-1}$-orbit are treated with the convolution of the calculated 
spectra with the wide width of $9.2$ MeV \cite{Tyren}. DWIA calculations 
of the $(K^-,K^+)$ cross sections are made as in the case of transitions 
to continuous states (TCS), except that the LSCA is not used to the 
$\Xi^-$ bound state wave functions. Figure \ref{fig-12CKmKp-Vdep-th} 
shows the calculated momentum spectra of $K^+$ at $p_{K^-}=1.65$ 
${\rm GeV}/c$ and $\theta_{K^+}=8^\circ$ in the region of $p_{K^+}$ 
where the peaks corresponding to the bound states appear. The calculated 
spectrum is convoluted by Lorentzian with the half width $\Gamma/2=2$ 
MeV which is the sum of $1$ MeV due to the decay 
$\Xi^-p\to\Lambda\Lambda$ and $1$ MeV due to the resolution of the 
detector in the experiment. 
\begin{table}[h]
\caption{\label{tab-EB}
Eigenstates and eigenenergies of $\Xi^-$ in the single-particle 
potential $U_{\Xi}$. 
}
\begin{tabular}{cccc} \hline \hline
 $V_{\Xi}$ [MeV] & \multicolumn{3}{c}{$E_{\Xi}$ [MeV]} \\ \cline{2-4}
 & 0$s$-state & 0$p$-state & 1$s$-state \\ \hline
 $-20$ &  $-8.3$ &  $-0.6$ & --- \\
 $-50$ & $-28.7$ & $-13.0$ & $-2.6$ \\ \hline \hline
\end{tabular}
\end{table}

In Fig. \ref{fig-12CKmKp-Vdep-th}, the dotted, solid, and dashed lines 
correspond to the calculated spectra with $V_{\Xi}=-50$, $-20$, and 
$+10$ MeV, respectively. One sees two (three) peaks corresponding to the 
transitions from the $0p_{3/2}^{-1}$-state to the $0s$- and $0p$-states 
($0s$-, $0p$- and $1s$-states) in Table \ref{tab-EB} for $V_{\Xi}=-20$ 
($-50$) MeV, while for $V_{\Xi}=+10$ MeV, as anticipated, there is no 
peak. 

An essential difference from the case of TCS is that there is no 
summation over the final states and, consequently, no short ranged 
kernel discussed in Sec. \ref{sec-2} that would restrict the distance of 
the source points of the outgoing distorted waves $\chi_f$ that can 
interfere with each other. Therefore, one needs the LSCA extrapolation 
of $\chi_f$ over a long distance, the validity of which needs to be 
examined. Numerical tests show that the accuracy of the LSCA 
extrapolation from a point on the nuclear surface to another point 
across the nuclear diameter is within about a factor of 2. This is the 
sort of accuracy of the calculations shown in 
Fig. \ref{fig-12CKmKp-Vdep-th}. 

The presence of a bound state peak in the spectrum shows that the 
$\Xi^-$-nucleus interaction is attractive and strong enough to support 
that particular bound state. Through the model calculation, one can 
estimate the strength of the peak that can suggest the experimental 
precision required for measuring it. In this regard, we consider what 
the results of the present work suggest is significant despite the 
numerical ambiguities discussed above. 

\section{\label{sec-5}SUMMARY AND CONCLUSIONS}
The inclusive $K^+$ momentum spectrum in the $^{12}$C$(K^-,K^+)$ 
reaction is calculated by the semiclassical distorted wave (SCDW) model. 
The same method of the calculation is used as in our previous works on 
$(\pi^-,K^+)$ and $(\pi^+,K^+)$ inclusive spectra \cite{Kohno, KFWOK}. 
The transition to the $\Xi^-$ bound state is also considered. The 
accuracy of the local semiclassical approximation (LSCA) for the 
discrete part of the spectrum is not so good as for the continuous part, 
but the error is estimated to be within a factor of 2. The standard 
$t\rho$ potentials are used for the $K^{\pm}$-$^{12}$C distorting 
potentials. Although the total cross sections of the $K^-$ and $K^+$ 
scatterings on $^{12}$C calculated with those potentials do not 
reproduce the experimental data very well, the resulting $(K^-,K^+)$ 
spectra agree with the ones calculated with the optical potentials 
$\hat{U}^{\pm}$ within about 10 \%. In order to improve a model for 
kaon distorted waves, one needs more experimental data in the momentum 
region of interest. 

Since we can explicitly take into account the dependences of the 
elementary $K^-+p\to K^++\Xi^-$ cross section on the invariant mass 
squared and the scattering angle, the present calculation is more 
realistic than those, e.g. in Ref. \cite{Tadokoro}, in which some 
constant average value of the elementary cross section is employed. The 
$\Xi^-$-nucleus potential is assumed to be of a Woods-Saxon form. The 
inclusive momentum spectra with the strength of the $^{12}$C$(K^-,K^+)$ 
reaction calculated with the strength of the potential $V_{\Xi}=-50$, 
$-20$, and $+10$ MeV are compared with the experimental data at 
$p_{K^-}=1.65$ ${\rm GeV}/c$. The shape of the spectrum is reproduced by 
the calculation. However, the magnitude of the cross section is 
underestimated by about a factor of 2. Though the inclusive spectrum 
changes systematically depending on the strength of $V_{\Xi}$, it is not 
possible to obtain a constraint on $V_{\Xi}$ from the present analysis 
on the basis of the available data. 

There are several possible reasons for the underestimation. There should 
be the contribution of the multi-step processes to $(K^-,K^+)$. We will 
investigate, in the next step, the influence of the multi-step processes 
by means of the SCDW model for these processes \cite{Weidenmuller}. The 
effect of in-medium modifications of the elementary process may not be 
negligible, although it can reduce the cross section. Another source of 
the discrepancy is experimental uncertainty of about 30 \% in the 
elementary cross sections used in the calculations. It is obvious that 
we need experimental data of the $(K^-,K^+)$ spectrum and the elementary 
cross sections with better accuracy for a more quantitatively reliable 
analysis. 

The calculated spectrum is found to have strong emission-angle 
dependence. When analyzing $(K^-,K^+)$ experimental spectra with 
angle-averaging involved, it is important to be aware of such 
dependence. It is seen from the spectra calculated at incident momenta 
of $p_{K^-}=1.50$, $1.65$, and $1.80$ ${\rm GeV}/c$ that the cross 
section at the lower side of the excitation energy hardly depends on the 
$K^-$ incident momentum. The close examination of the calculation shows 
that this consequence is due to the Fermi motion of the target nucleons, 
which is properly treated in our SCDW framework. The discrete part of 
the spectrum of $K^+$ will be useful for extracting information on the 
$\Xi^-$-nucleus potential from the future $(K^-,K^+)$ experimental data, 
because the possibility to detect a peak structure below the threshold 
has not been ruled out yet experimentally. 

\begin{acknowledgments}
The authors would like to thank T. Fukuda and A. Ohnishi for valuable 
discussions. This study is supported by Grants-in-Aid for Scientific 
Research from the Japan Society for the Promotion of Science 
(Grant Nos. 18042004 and 17540263). 
\end{acknowledgments}

\bibliography{basename of .bib file}



\end{document}